\documentclass[aps,prb,reprint,groupedaddress]{revtex4-1}
\usepackage{graphicx}
\usepackage{csquotes}

\begin{document}

\title{Frequency-dependent Study of Solid Helium-4 Contained in a Rigid Double-torus Torsional Oscillator}


\author{Jaewon Choi}
\email[]{rappinmind@kaist.ac.kr}
\author{Jaeho Shin}
\author{Eunseong Kim}
\email[Corresponding author: ]{eunseong@kaist.edu}
\affiliation{Department of Physics and Center for Supersolid and Quantum Matter Research, Korea Advanced Institute of Science and Technology (KAIST), 291 Daehak-ro, Yuseong-gu, Daejeon 34141, Republic of Korea}

\date{\today}

\begin{abstract}
The rigid double-torus torsional oscillator (TO) is constructed to reduce any elastic effects inherent to complicate TO structures, allowing explicit probing for a genuine supersolid signature. We investigated the frequency- and temperature-dependent response of the rigid double-torus TO containing solid 4He with 0.6 ppb 3He and 300 ppb 3He. We did not find evidence to support the frequency-independent contribution proposed to be a property of supersolid helium. The frequency-dependent contribution which comes from the simple elastic effect of solid helium coupled to TO is essentially responsible for the entire response. The magnitude of the period drop is linearly proportional to ${f}^{2}$, indicating that the responses observed in this TO are mostly caused by the overshoot of \enquote*{soft} solid helium against the wall of the torus. Dissipation of the rigid TO is vastly suppressed compared to those of non-rigid TOs.
\end{abstract}

\pacs{67.80.bd}
\pacs{67.80.de}
\keywords{Supersolid, Solid Helium, Torsional Oscillator}

\maketitle


\section{introduction}
The resonant period of an ideal torsional oscillator (TO) is proportional to the square root of its rotational inertia $\sqrt{I}$, and the superfluid decoupling can be detected by the reduction of the resonant period. The resonant period drop of a TO containing solid helium was originally interpreted as the appearance of a supersolid phase \cite{kim2004probable, kim2004observation, andreev1969quantum, chester1970speculations, penrose1956bose, thouless1969flow, leggett1970can, kondo2007observation, aoki2007oscillation, rittner2006observation, rittner2007disorder, penzev2008ac, hunt2009evidence, choi2010evidence, choi2010observation, zmeev2011simultaneous, kim2011unaffected}. Recently, a number of experimental and theoretical efforts have indicated that the anomaly in the TO response can be explained by the shear modulus change \cite{day2007low, day2009intrinsic, day2010nonlinear, beamish2010frequency, reppy2010nonsuperfluid, fefferman20124,  haziot2013giant, haziot2013dislocation, haziot2013critical, fefferman2014dislocation}. The previous finite element method (FEM) simulation suggested that the influence on the period of TO due to the change in shear modulus of solid helium was negligible \cite{clark2008thermal, maris2011effect, reppy2012interpreting}. Nevertheless, the effect can be significantly amplified resulting from non-ideal TO design. Four mechanisms of non-ideal TO response have been suggested \cite{reppy2012interpreting, beamish2012elastic, maris2012effect, kim2012absence}. In order to minimize the contribution of the elastic effect to the resonant period of TO, it should be meticulously constructed to be rigid. However, most of TOs used in the previous supersolid experiments were not rigid. \cite{reppy2012interpreting, beamish2012elastic, maris2012effect}. Recently, the Chan group reported that the resonant period drop was substantially reduced when the rigidity of TO was systematically increased \cite{west2009role, kim2012probing, kim2014upper}. The resonant period drop of a highly rigid TO was only a few times greater than that due to the elastic effect estimated by FEM simulation. They set the upper bound for the non-classical rotational inertia (NCRI) to be less than approximately 4 ppm \cite{kim2014upper}.

However, new evidence of a \enquote*{true} supersolid signature was suggested by Reppy et al. \cite{mi2014pursuit, mi2014evidence} based on double-frequency TO experiments. Superfluidity or the NCRI is independent of frequency while non-superfluidity or the relaxation phenomenon resulting from shear modulus change leads to a strong frequency dependence. Accordingly, analysis on the frequency dependence can be used to differentiate the superfluid response from the non-superfluid response \cite{reppy2012interpreting, mi2014evidence}. Reppy et al. observed a small frequency-independent resonant period change after subtracting the frequency-dependent term and ascribed this to a possible supersolid signature. This interpretation can be questioned since the measurements reported relatively large period drop which can be associated with the non-rigidity of TO.

For this article, we measured the period drop and accompanied dissipation peak using a double-frequency TO that was constructed to be highly rigid to minimize shear modulus effects. The frequency dependence of the rigid double-frequency TO was investigated in various modes of representation. As a result, we elucidate whether or not solid helium-4 exhibits true supersolidity.

\section{experimental details}

\begin{figure}
\includegraphics[width=0.5\textwidth]{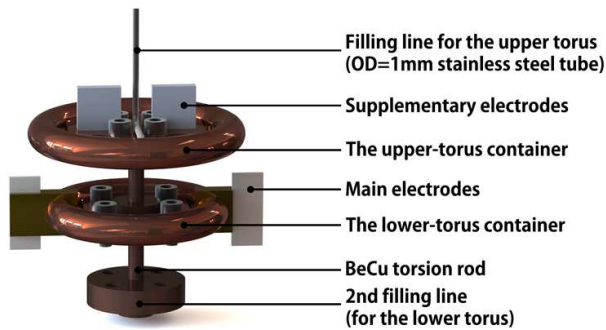}
\caption{KAIST rigid double-torus torsional oscillator.\label{fig:01}}
\end{figure}

KAIST rigid double-torus torsional oscillator (TO) (Fig. \ref{fig:01}) was carefully designed to minimize the various elastic effects caused by: (1) shear-modulus-dependent relative motion between TO parts (the glue effect) \cite{reppy2012interpreting, kim2012absence}, (2) solid helium contained in the torsional rod (the torsion rod hole effect) \cite{beamish2012elastic}, and (3) solid helium layer grown on a thin TO base plate (Maris effect) \cite{maris2012effect}.

We first assembled every joint in the TO using stainless-steel screws or hard-soldering to diminish elastic effects on the TO period due to the relative motion between different parts. The torsion plate and the torsion rod were rigidly connected by machining the assembly from a single piece of Be-Cu. The torus-shaped TO cell for solid helium was constructed by hard-soldering two pieces of semi-circular copper tubing. The torus was hard-soldered onto thick copper plate and the combined structures were fastened down directly on the torsion plate by four screws. 

Second, we used a thick torsion rod and a very thin fill line to prevent the elastic effect of solid helium in the fill line. The change of shear modulus of solid helium-4 contained in the torsion rod can induce the TO period anomaly \cite{beamish2012elastic}. This effect may be responsible for the period drop observed in the majority of TO experiments. In order to remove this effect, a 1-mm-diatmeter CuNi filling capillary was directly installed on top of the TO cell instead of making a hole through a torsion rod.

Third, the TO cell containing solid helium was hard-soldered to a 3-mm-thick copper plate and was designed so that solid helium in the torus did not have a direct contact to the base plate near the torsion rod. In a cylindrical TO, solid helium was grown directly on its base plate. If the thickness of the base plate is not sufficiently thick, then the period drop due to the change in shear modulus of solid helium can be significantly amplified \cite{maris2012effect}. The contribution of solid helium is greater in the proximity of the torsion rod and when the base plate is thinner. In this study, solid helium confined in a rigid torus channel on the thick copper plate was not expected to exhibit the strong Maris effect. 

Finally, we carefully tuned the so-called \enquote*{overshoot effect} caused by direct coupling of elastic properties of solid helium to the TO response. Since solid helium is much softer than the wall of the container, it undergoes additional displacement. The shear modulus change of solid helium not moving in phase with the confining wall can induce the period anomaly without non-linear amplification. The resonant period drop introduced by the overshoot effect is known to be linearly proportional to the square of the TO frequency. For this experiment, we optimized the overshoot effect so that the system was in a regime where the elastic effect is not too small to be detected and not too big to make the TO softer. In this TO, solid helium was located in a torus with a relatively large cross-section (diameter: 5 mm), which allows us to have large mass loading and high resolution for the so-called NCRI fraction. One can effectively reduce this effect by confining solid helium in a thick copper torus with a small cross-section at the expense of diminishing the capability of frequency analysis. This fine-tuning allows us to investigate the effect of the shear modulus on the double-frequency TO and possible supersolidity more clearly.	

In addition, the KAIST rigid double-torus TO can be operated at four resonant frequencies. This is possible due to the configuration of our TO which consists of two torus-shaped solid helium containers attached to upper and lower stages. This configuration enables modification of the resonant frequency by loading solid helium into the upper or both tori. We first placed the solid sample in both tori. After collecting the first dataset, the solid helium inside the lower torus was carefully removed at low temperature to avoid damaging the solid sample placed in the upper torus.

Finite element method (FEM) simulations indicates that a 30\% change in the shear modulus of solid helium-4 in the upper torus leads to only a 0.93-ns period drop in first (1st) mode (in-phase) and a 1.16-ns drop in the second (2nd) mode (out-of-phase). These simulations correspond to $2.3 \times 10^{-5}$ and $1.9 \times 10^{-4}$ respectively in the framework of so-called NCRI fraction. The major contribution seemingly comes from the overshoot effect, the relative motion of solid helium with respect to the outer wall. 

The empty cell has resonant frequencies of 449.57 Hz and 1139.77 Hz for in-phase and out-of-phase modes respectively. Two pairs of electrodes attached to the lower torus were used to drive and detect the TO response. The additional electrodes installed on the upper plate enables the detection of the amplitude and phase of the upper torus. We confirmed that the phase difference between the two tori was approximately 0 degrees for the in-phase mode and 180 degrees for out-of-phase mode. The mechanical quality factor was approximately $1 \times 10^{6}$ at 4.2 K for both modes. 

Bulk solid helium-4 samples containing an impurity concentration of 0.6 ppb and 300 ppb helium-3 were grown by the blocked capillary method. The sample cell was first pressurized to target pressures of 67-82 bar at 3.2 K and the mixing chamber was then cooled to the base temperature. The resonant period shifted according to solid growth in the upper torus: 39,800 ns for the in-phase mode and 6,040 ns for the out-of-phase mode. The large period shift due to solid helium and its solid stability in the first mode enables us to distinguish the change of the resonant period within about 2 ppm resolution. This sensitivity is lower than the upper limit of NCRI fraction reported by the PSU group (4 ppm) \cite{kim2014upper}.

\section{TO period and dissipation}

\begin{figure*}
\includegraphics[width=1\textwidth]{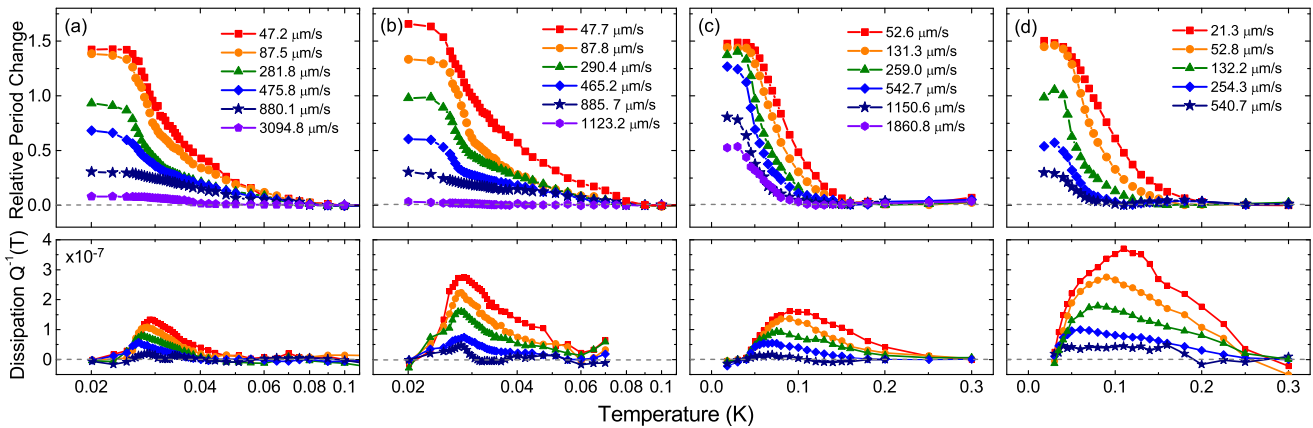}
\caption{Period and dissipation data of the KAIST rigid double-pendulum TO containing solid helium with 0.6-ppb 3He ((a) in-phase and (b) out-of-phase) and 300-ppb 3He ((c) in-phase and (d) out-of-phase). The empty-cell background was subtracted in both period and dissipation data. The y-axis of upper panels represent absolute values of period changes, calculated from the empty-cell backgrounds. \label{fig:02}}
\end{figure*}


The period and dissipation of the TO for both the in-phase and out-of-phase modes were measured over a temperature range of 20-300 mK. Figure \ref{fig:02} presents the resonant period of the rigid double-torus TO containing high-purity solid helium-4 (0.6 ppb) and commercial-purity solid helium-4 (300 ppb) as a function of temperatures at different rim velocities. The period drop anomaly was observed for both modes of the rigid double-torus TO.

Compared with the empty cell data (dashed line), the period drop in 0.6-ppb solid sample appears initially at 80 mK, and became saturated below 25 mK for both modes (colored symbols). The magnitude of the period drop is suppressed by the increase in the rim velocity of the alternating current (AC) oscillation for both modes. The suppression of the TO anomaly appears at a rim velocity of ~100 $\mu$m/s for both modes. Similar behaviors have been observed in numerous previous experiments \cite{kim2004probable, kim2004observation, kondo2007observation, aoki2007oscillation, rittner2006observation, rittner2007disorder, penzev2008ac, hunt2009evidence, choi2010evidence, choi2010observation}, including the measurement of the rigid TOs \cite{west2009role, kim2012probing, kim2014upper}. The low onset temperature of about 80 mK and sharper temperature dependence of the TO anomaly were reported for low helium-3 impurity concentrations \cite{day2007low, kim2008effect}. The period values measured at different rim velocities follows the empty cell background at high temperatures for both resonant modes.

The maximum period drop at the lowest rim velocity is 1.43 ns and 1.66 ns for the in-phase (-) and the out-of-phase (+) modes respectively, which are a similar order of magnitude to those expected from the contribution by the shear modulus (0.93 ns for the in-phase mode and 1.16 ns for the out-of-phase mode with a 30\% shear modulus change). By subtracting the empty-cell background, we calculated ${dP}_{\pm}$ with respect to the mass loading of solid helium ${\Delta P}_{\pm}$, equivalent to the NCRI fraction, ${dP}_{-}/{\Delta P}_{-}=3.5 \times 10^{-5}$ for the in-phase mode; $dP_{+}/\Delta P_{+}=2.5 \times 10^{-4}$ for the out-of-phase mode. Considering that the maximum change in shear modulus at the lowest temperature can vary from 8\% \cite{day2007low} to 86\% \cite{rojas2010anomalous}, the period drop can be reasonably attributed to the stiffening effect of the shear modulus in solid helium. Dissipation in the TO response is also observed for both modes. The dissipation peak appears around 30 mK at which the period of the TO changes most drastically. The dissipation peak is also suppressed by increasing the rim velocity in both modes agreeing with previous measurements \cite{kim2004probable, kim2004observation, kondo2007observation, aoki2007oscillation, rittner2006observation, rittner2007disorder, penzev2008ac, hunt2009evidence, choi2010evidence, choi2010observation}.

We observed essentially the same results for the commercial-purity (300 ppb) solid helium-4 sample except for the anomaly found at higher temperatures. In the in-phase mode, the anomalous period drop appears at an onset temperature of 160 mK and reaches a maximum of 1.49 ns ($dP_{-}/\Delta P_{-} = 3.7 \times 10^{-5}$) at 40 mK. In the out-of-phase mode, the onset temperature is higher than that of the in-phase mode, about 200 mK. The shifted onset temperature at higher frequency modes has been previously reported from a double-pendulum TO by Rutgers group \cite{aoki2007oscillation} and also in shear modulus measurements in another experiment \cite{beamish2010frequency}. The magnitude of the period drop saturates to 1.55 ns ($dP_{-}/\Delta P_{-} = 2.6 \times 10^{-4}$) at 40 mK. The maximum period drop observed for both modes has a similar order of magnitude as those anticipated by the FEM simulation. The dissipation in TO amplitude appears over the same temperature range. At the lowest rim velocity of 50 $\mu$m/s, the dissipation peak was located at approximately 105 mK.

We investigated seven different solid helium samples. The resonant period showed similar order of magnitude drops, approximately 1.3-1.7 ns for both modes. However, only two solid samples described above show dissipation in the TO response. In addition, the size of these dissipation peaks is approximately $10^{-7}$ and is orders of magnitude smaller than those from typical TO experiments. These minute dissipation features are consistent with the results from the Chan group who showed no clear dissipation features in their rigid TO experiments \cite{kim2012probing, kim2014upper}. The absence of and/or the significant reduction of dissipations can be connected to a rigidity of the TO.

\section{frequency-dependent study}

The frequency dependence of the TO responses is examined to clarify the origin of the marginal period drop. While $dP/\Delta P$ is independent of the measurement frequency in the supersolid scenario, it is proportional to the square of the measurement frequency in the shear-modulus effect scenario \cite{reppy2012interpreting}. Reppy et al. \cite{mi2014pursuit, mi2014evidence} provided a simple mathematical methods to decompose the period drop observed in their TO experiment into both frequency-independent and frequency-dependent parts explicitly. The measured period drop consists of two terms: (i) the frequency-independent term ${[{dP}_{\pm}/{\Delta P}_{\pm}]}^{ind}(T,V)$, regarded as a putative supersolid signature, and (ii) the frequency-dependent term ${[{dP}_{\pm}/{\Delta P}_{\pm}]}^{dep}(T,V,f)$, attributed to the elastic overshoot effect, where T is temperature and V is rim velocity. The total period drop observed in TO experiments can be written as follows:

\begin{equation} \label{eq:eps}
{\left[ \frac{{dP}_{\pm}}{{\Delta P}_{\pm}}\right]}^{exp}={\left[ \frac{{dP}_{\pm}}{{\Delta P}_{\pm}}\right]}^{ind}(T,V)+{\left[ \frac{{dP}_{\pm}}{{\Delta P}_{\pm}}\right]}^{dep}(T,V,f)
\end{equation}

Since the period drop originating from the overshoot effect is proportional to ${f}^{2}$, the last term can be substituted with ${[{dP}_{\pm}/{\Delta P}_{\pm}]}^{dep}(T,V,f)=a(T,V){f}^{2}$. Then, both the frequency-independent term ${[{dP}_{-}/{\Delta P}_{-}]}^{ind}$ and the frequency-dependent term ${[{dP}_{-}/{\Delta P}_{-}]}^{dep}$ of the in-phase mode are decomposed as follows:

\begin{equation} \label{eq:eps}
{\left[ \frac{{dP}_{-}}{{\Delta P}_{-}}\right]}^{ind}= \frac{{f}_{+}^{2}}{\gamma}{\left[ \frac{{dP}_{-}}{{\Delta P}_{-}}\right]}^{exp}-\frac{{f}_{-}^{2}}{\gamma}{\left[ \frac{{dP}_{+}}{{\Delta P}_{+}}\right]}^{exp} 
\end{equation}

\begin{equation} \label{eq:eps}
{\left[ \frac{{dP}_{-}}{{\Delta P}_{-}}\right]}^{dep}= \left(1-\frac{{f}_{+}^{2}}{\gamma}\right){\left[ \frac{{dP}_{-}}{{\Delta P}_{-}}\right]}^{exp}+\frac{{f}_{-}^{2}}{\gamma}{\left[ \frac{{dP}_{+}}{{\Delta P}_{+}}\right]}^{exp} 
\end{equation}
where $\gamma={f}_{+}^{2}-{f}_{-}^{2}$.

\begin{figure}
\includegraphics[width=0.5\textwidth]{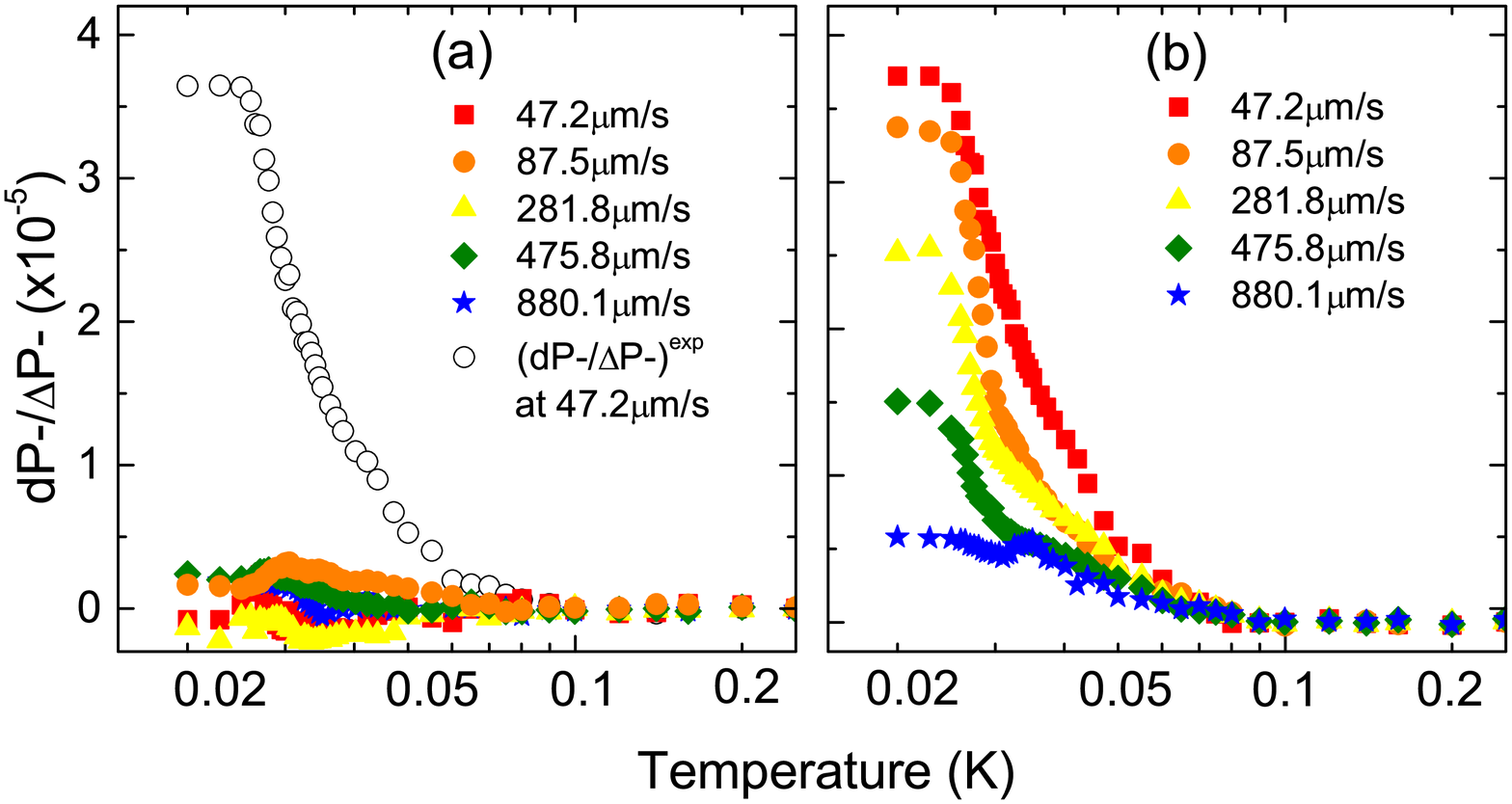}
\caption{(a) Frequency-independent ${{dP}_{-}/{\Delta P}_{-}}^{ind}$ and (b) frequency-dependent ${{dP}_{-}/{\Delta P}_{-}}^{dep}$ as a function of temperature. The same color-coded pair for each mode was obtained using the same rim velocity. No distinguishable velocity effect is seen in ${{dP}_{-}/{\Delta P}_{-}}^{ind}$ (a), in contrast to (b) demonstrating a clear rim-velocity dependence. ${{dP}_{-}/{\Delta P}_{-}}$ at the lowest rim velocity 47.2$\mu$m/s is shown for comparison.\label{fig:03}}
\end{figure}

To measure ${[{dP}_{-}/{\Delta P}_{-}]}^{ind}$ and ${[{dP}_{-}/{\Delta P}_{-}]}^{dep}$ at a certain rim velocity, the driving AC voltage was carefully adjusted to have the same rim velocity for the in-phase and out-of-phase modes. Accordingly, the same color-coded pair of temperature scans for each mode as shown in Figure \ref{fig:03} was obtained at the same rim velocity, despite of different driving AC voltages. 

The temperature dependence of (a) the frequency-independent and (b) the frequency-dependent term at various TO rim velocities are plotted in Figure \ref{fig:03}. The datasets identified with the same color in both Figure \ref{fig:03}-(a) and \ref{fig:03}-(b) are obtained using the same rim velocity. However, the two figures are significantly different. No sizable frequency-independent period drop at any rim velocity can be identified in Figure \ref{fig:03}-(a), while the frequency-dependent period response at the rim velocity of 47.2 $\mu$m/s is nearly the same as the total period drop in the measurements. The averaged frequency-independent period drop is approximately 0$\pm$4 ppm, which is similar to the upper limit set in the rigid TO study \cite{kim2014upper}. Accordingly, the majority of the period anomaly can be attributed not to the appearance of supersolidity but to the stiffening of the shear modulus of solid helium-4. Besides, the frequency-dependent term was strongly suppressed with increasing TO rim velocity. In contrast, no apparent drive dependence was observed for the frequency-independent drop. The frequency-dependent TO response can be extrapolated to values less than 4 ppm (0.16 ns) when the stiffening of solid helium is significantly suppressed, indicating that the entire TO anomaly can be ascribed to the shear modulus change of solid helium at low temperatures. 

\begin{figure}
\includegraphics[width=0.5\textwidth]{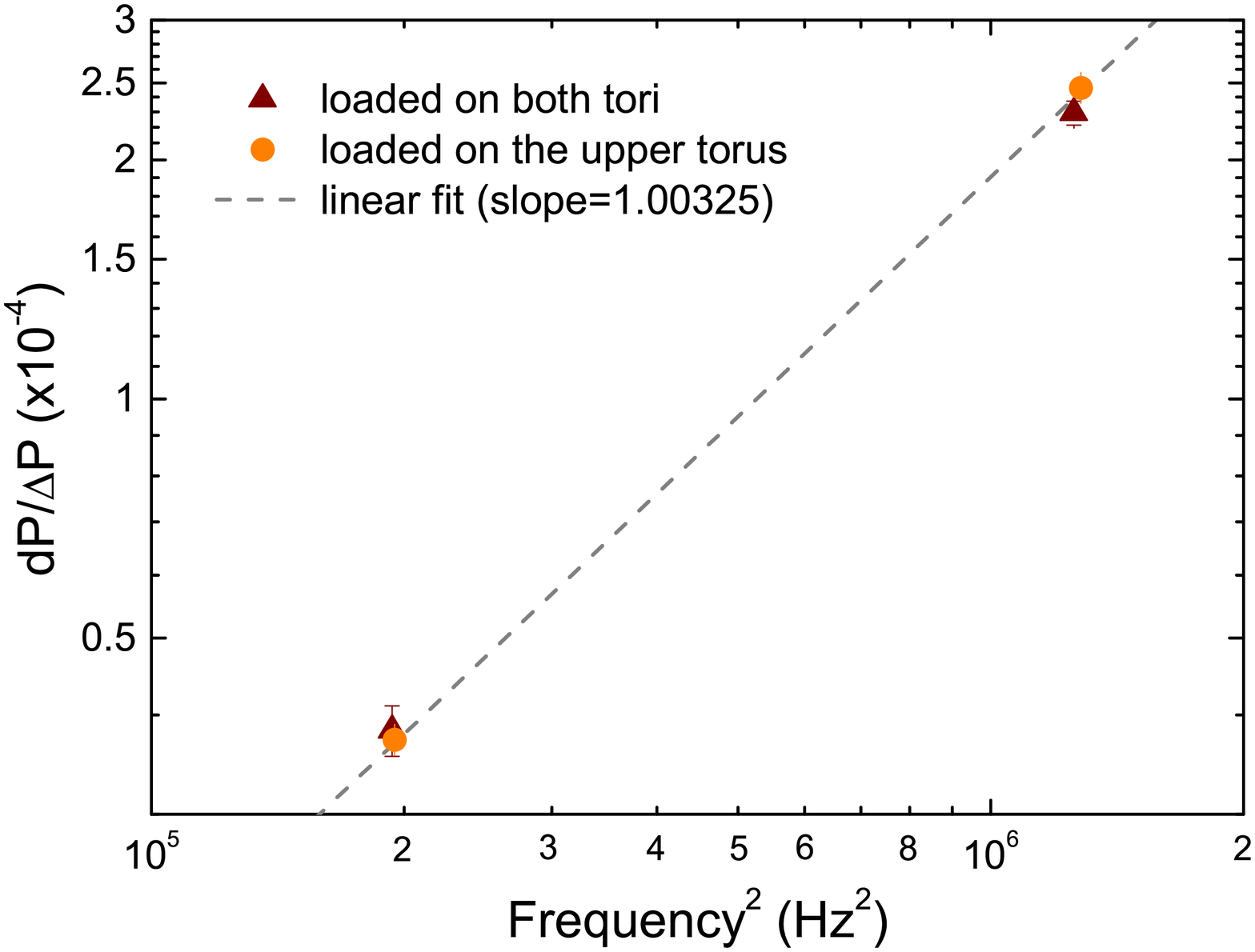}
\caption{Log-Log plot of $dP/\Delta P$ measured in four different frequencies as a function of the TO frequency squared for 0.6-ppb 3He (solid triangle) and 300-ppb 3He (solid circles). The red solid triangles represent $dP/\Delta P$ measured with both tori filled with the solid helium sample, while the orange solid circles represent $dP/\Delta P$ measured with only upper torus filled. The grey dashed line indicates the linearly fitted result. The y-intercept can be converted to frequency-independent $dP/\Delta P$ of -0.3ppm. \label{fig:04}}
\end{figure}

Figure \ref{fig:04} shows the low temperature $dP/\Delta P$ for four resonant modes as a function of frequency. Two solid triangles are measured for solid samples grown in both tori and the other two solid circles corresponded to samples grown only in the upper torus. In the log-log plot of $dP/\Delta P$ and ${f}^{2}$, the data can be linearly fitted to the equation $\log \left(dP/\Delta P \right)=(-9.74)+(1.003)\log({f}^{2})$. The slope of equation is nearly 1, indicating that $dP/\Delta P$ is linearly proportional to ${f}^{2}$. Converting the log-scale axis to a linear one, the y-intercept is equivalent to $-2.62 \times 10^{-7}$ (-0.3 ppm) which indicates that the measured period drop originates from the shear modulus effect rather than supersolid mass decoupling.

The frequency dependence can be analyzed by other methods. The ratio between the period shifts of the two modes ${\delta P}_{+}/{\delta P}_{-}$ can distinguish the origin of TO response at low temperature: either supersolidity or shear modulus change \cite{reppy2012interpreting}. The ratio for the supersolid scenario ${\left({\delta P}_{+}/{\delta P}_{-}\right)}_{SS}$ would follow the mass-loading (or missing) change which can be easily obtained by measuring the mass-loading-induced period shifts of both modes. In contrast, the ratio for the shear modulus effect required an additional frequency-dependent contribution as follows:

\begin{equation} \label{eq:eps}
{\left({\delta P}_{+}/{\delta P}_{-}\right)}_{SM}=\left({f}_{+}^{2}/{f}_{-}^{2}\right){\left({\delta P}_{+}/{\delta P}_{-}\right)}_{SS}
\end{equation}

\begin{figure}
\includegraphics[width=0.5\textwidth]{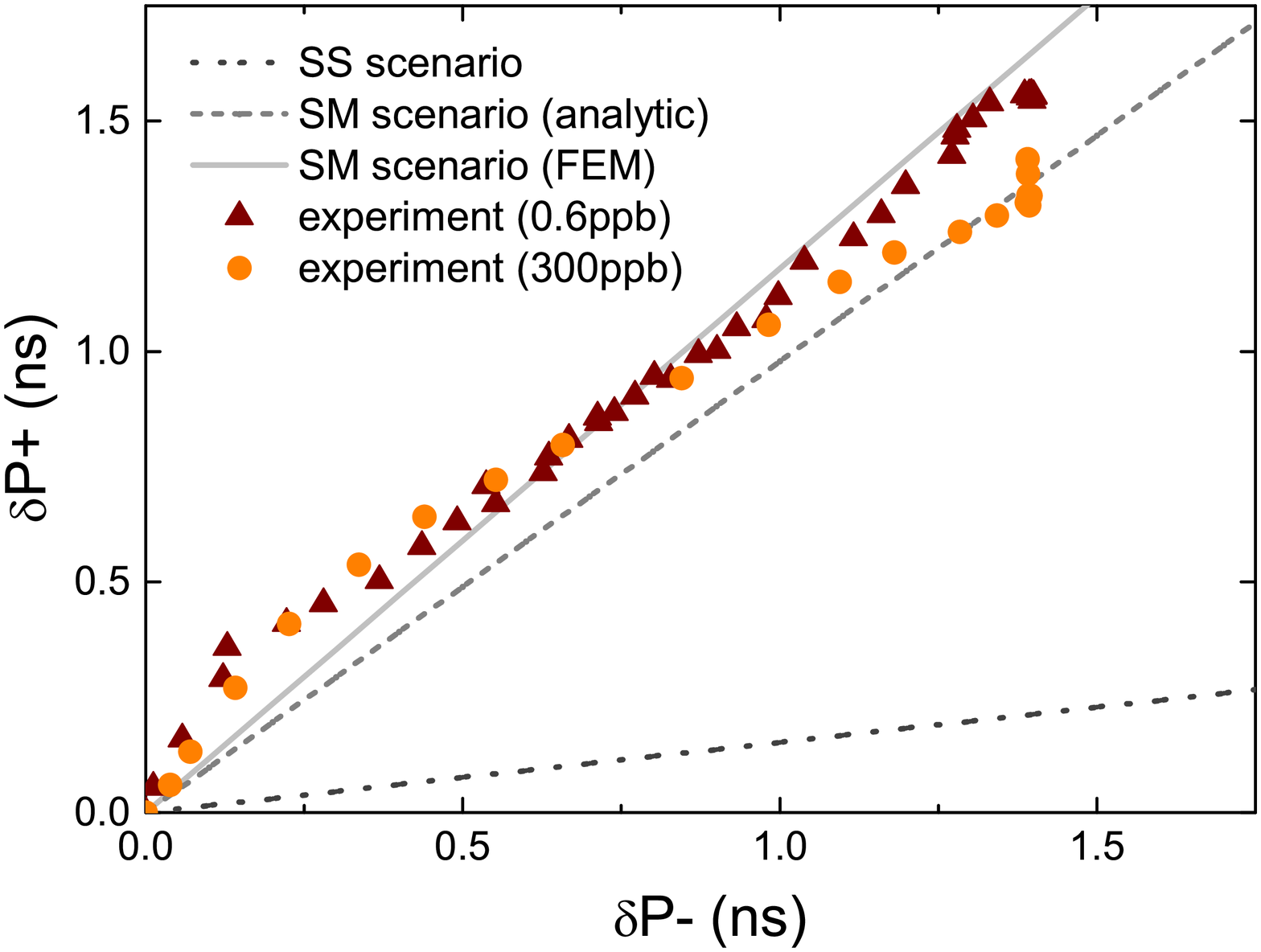}
\caption{Plot of the ratio of the period shifts measured in both modes. The dotted line, the supersolid mass decoupling (SS) scenario, is directly calculated by the ratio of mass-loading in both modes. Both solid and dashed lines indicate the shear modulus (SM) scenario estimated by using finite element method (FEM) simulation and an analytical solution, respectively. The experimental results of solid helium with 0.6-ppb 3He (solid triangles) and 300-ppb 3He (solid circles) are consistent with the SM scenario.  \label{fig:05}}
\end{figure}

The mass-loading ratio ${\left({\delta P}_{+}/{\delta P}_{-}\right)}_{SS}$ in our experiment is measured to be approximately 0.152. The solid straight line shown in Figure \ref{fig:05} indicates mass decoupling or the supersolid scenario estimated by FEM simulation and experimental measurements. The effect from the shear modulus change of solid helium is estimated analytically (the dashed line) and with FEM simulation (the solid line). The slope of ${\delta P}_{+}/{\delta P}_{-}$ plot in the shear modulus scenario is steeper than that in the supersolid mass-loading scenario due to the additional frequency-dependent contribution. The measured ${\delta P}_{+}/{\delta P}_{-}$ values for both 0.6 ppb and 300 ppb solid helium-4 (Figure \ref{fig:05}) indicate that the ratio ${\delta P}_{+}/{\delta P}_{-}$ collapses to the shear modulus scenario. The ratios obtained in the previous studies lie between shear modulus expectations and supersolid expectations, suggesting the possible existence of a putative supersolid \cite{aoki2007oscillation, mi2014pursuit, mi2014evidence, nichols2013frequency}. However, we confirmed that the TO responses from the KAIST rigid double-torus TO originated from the non-supersolid origin. The discrepancy with previous double-frequency TO observations may arise from the rigidity of TO.

\section{discussions}

Recently, Both the Cornell \cite{mi2014pursuit, mi2014evidence} and London \cite{nichols2013frequency} groups constructed a double-frequency TO to investigate the frequency dependence of the period anomaly. The London group measured the period and dissipation of a two-mode TO containing a poly-crystalline solid helium-4 sample. They observed a period drop and concomitant dissipation features, equivalent to ${dP}_{-}/{\Delta P}_{-}=2.10 \times 10^{-3}$ and ${dP}_{+}/{\Delta P}_{+}=8.04 \times 10^{-3}$. The torsion rod hole effect and Maris effect was removed by analytical calculations. After fitting a linear equation to those data in ${f}^{2}$-domain, they found a sizable frequency-independent period drop ${[{dP}_{-}/{\Delta P}_{-}]}^{ind}=1.86 \times 10^{-3}$. The period drop and dissipation were analyzed by the complex response function. The TO response they observed was not in agreement with the functional form of simple glassy dynamics. The authors concluded that a different physical mechanism is required for explaining the TO responses.

The Cornell group also designed a compound TO with an annular sample space and measured the TO responses. After removing the overshoot effect by frequency analysis, they extracted a finite frequency-independent period drop ${[{dP}_{-}/{\Delta P}_{-}]}^{ind}=1 \times 10^{-4}$, several orders of magnitude larger than the elastic contribution estimated by their FEM simulation. Additional dissipation introduced by solid helium was measured to be very small. The height of the dissipation peak was reported to only $5 \times 10^{-8}$ in both resonant modes. They proposed that the finite frequency-independent period drop could possible be new evidence for supersolidity in bulk solid helium.

In our rigid TO, the frequency-independent period drop is two or three orders of magnitude smaller than the value reported from the Cornell and London group. The dissipation peak is not observed in most solid samples or the size of the dissipation peak was very small ($\sim{10}^{-7}$). We believe that this discrepancy is presumably due to TO rigidity. If the ‘true’ NCRI fraction is about 100 ppm, then the period drop anomaly should have been observed in highly rigid TO experiments at PSU and in our rigid double TO experiments. 

\section{conclusion}

We studied the frequency dependence of TO responses of solid helium-4 using the rigid double-torus torsional oscillator. The period drop anomaly is observed for both in-phase and out-of-phase modes. Frequency analysis shows that the frequency-independent period shift is less than 4 ppm, close to the upper limit set by the PSU group, and the frequency-dependent contribution is almost the same as the TO response. We conclude that the TO response at low temperatures is not due to the appearance of supersolidity but due to the change in the shear modulus of solid helium. The supersolid fraction, if it exists, should be smaller than 4 ppm.

\begin{acknowledgments}
The authors acknowledge Duk Y. Kim at Pennsylvania State University for fruitful discussions and advices in the early stage of construction of the rigid double-pendulum torsional oscillator. This work is supported by the National Research Foundation of Korea through the Creative Research Initiatives. J. Choi would like to thank the POSCO TJ Park Foundation for its financial support and generosity through the TJ Park Science Fellowship.
\end{acknowledgments}

\bibliography{2015Choi_PRB}

\end{document}